\documentclass[preprint]{aastex631}

\usepackage{color}
\usepackage{graphicx}
\shortauthors{Shi et al.}

\begin{document}

\title{Constraining baryon loading efficiency of AGNs with diffuse neutrino flux from galaxy clusters}

\author[0000-0003-1244-172X]{Xin-Yue Shi}
\affil{Department of Astronomy, Nanjing University, 163 Xianlin Avenue, Nanjing 210023, China}
\affil{Key Laboratory of Modern Astronomy and Astrophysics, Nanjing University, Ministry of Education, Nanjing, China}

\author[0000-0003-1576-0961]{Ruo-Yu Liu}
\affil{Department of Astronomy, Nanjing University, 163 Xianlin Avenue, Nanjing 210023, China}
\affil{Key Laboratory of Modern Astronomy and Astrophysics, Nanjing University, Ministry of Education, Nanjing, China}

\author[0000-0003-0628-5118]{Chong Ge}
\affil{Department of Astronomy, Xiamen University, Xiamen, China}

\author[0000-0002-5881-335X]{Xiang-Yu Wang}
\affil{Department of Astronomy, Nanjing University, 163 Xianlin Avenue, Nanjing 210023, China}
\affil{Key Laboratory of Modern Astronomy and Astrophysics, Nanjing University, Ministry of Education, Nanjing, China}

\correspondingauthor{Ruo-Yu Liu}
\email{ryliu@nju.edu.cn}

\begin{abstract}
The active galactic nuclei (AGNs) are widely believed to be one of the promising acceleration sites of ultrahigh-energy cosmic rays (CRs). Essentially, AGNs are powered by the gravitational energy of matter falling to supermassive black holes. However, the conversion efficiency of gravitational to kinetic energy of CRs in AGNs, which is defined as baryon loading factor $\eta_p$, is not well known yet.
After being accelerated, high-energy CRs could escape the host galaxy and enter the intra-cluster medium (ICM). 
These CRs can be confined within the galaxy cluster and produce $\gamma$-rays and neutrinos through proton-proton collisions with the ICM.
In this paper, we study the diffusion of CRs in galaxy clusters and calculate the diffuse neutrino flux from galaxy cluster population.
Using the latest upper limits on the cumulative unresolved TeV-PeV neutrino flux from galaxy clusters posed by the IceCube Neutrino Observatory, we derive the upper limit of the average baryon loading factor as $\eta_{p,\mathrm{grav}} \lesssim 2 \times 10^{-3} - 0.1$ for the population of galaxy clusters.
This constraint is more stringent than the one obtained from $\gamma$-ray observation on the Coma cluster.
\end{abstract}


\section{Introduction} \label{sec:intro}

Active galactic nuclei (AGNs) are most powerful persistent emitters of radiation in the universe and have been consider as potential source of extragalactic high-energy cosmic rays (CRs) \citep{Biermann1988, Takahara1990, Rachen1993, Berezinsky2006, Dermer2009} and neutrinos\citep{Mannheim1992, Stecker1996, Atoyan2001, Murase2014}.
Acceleration of baryonic CRs in jets/outflows of AGNs consumes the kinetic energy or magnetic energy of the jets/outflows, which are essentially fueled by the gravitational energy of matter falling into supermassive black hole at center of the nuclei.
The efficiency of gravitational energy converting into CRs, which is defined as the baryon loading factor $\eta_p$, can help us understand the physical mechanism of particle acceleration.

The IceCube Neutrino Observatory has been observing TeV–PeV astrophysical neutrinos for over one decade. All the discovered potential sources are associated with AGNs,  such as 
the blazar TXS 0506+056 \citep{IceCube2018a,IceCube2018b} and the Seyfert II galaxy NGC 1068 \citep{Aartsen2020} detected by IceCube. Indeed, numerous studies have shown that protons accelerated in AGNs can interact with their intense radiation fields and produce high-energy neutrinos via the photohadronic interactions \citep{Rachen1998, Atoyan2003, Stecker2013} or hadronuclear interactions\citep{Fraija2012, Sahakyan2013, Li2022, Xue2022}. On the other hand, whether the all-sky diffuse neutrino background can be accounted for by the AGN population highly depends on the average baryon loading factor of the AGN population, because the expected neutrino flux is proportional to this parameter. For example, \citet{Murase2014} found that
blazars may account for the diffuse neutrino background above 100\,TeV with $\eta_{p,\mathrm{rad}} \equiv L_p/L_\gamma \sim 3-300$ assuming a $E_p^{-2}$ CR proton spectrum.
However, the baryon loading mechanism of AGN jets or outflows is not well known yet, which prevents us from drawing a concrete conclusion on the contribution of AGNs to the all-sky diffuse neutrino background  \citep{Berezhko2008, Cuoco2008, Kadler2016, Righi2017, Palladino2019}.

On the other hand, CRs generally lose only a small fraction of energies through the hadronic interactions in AGNs \citep[e.g.,][]{Murase2014, Xue2019}. Thus, they may eventually escape AGNs and host galaxies after being accelerated, and propagate into the intra-cluster (ICM) of the galaxy cluster. Galaxy clusters are the largest gravitational bound structures in the universe. They are also known as efficient reservoirs for CRs (\citealt{Voelk1996, Berezinsky1997}, see \citealt{Brunetti2015} for a recent review). The diffusion timescale of CRs with energies $\lesssim 1$ PeV are longer than the Hubble timescale with reasonable diffusion coefficient in the ICM.
The confined CRs would interact with the ICM via $pp$ collision and produce $\gamma$-rays and neutrinos. Therefore, measurements on $\gamma$-rays and neutrinos from galaxy clusters can serve as constraints on the amount of CRs accelerated in AGNs , which can be translated to the baryon loading factor.

The $\gamma$-ray emissions from individual galaxy cluster has been searched in very-high-energy band ($\textgreater$ 100 GeV) by Fermi-LAT\citep{Han2012, Ackermann2016} and Imaging Air Cherenkov Telescopes (IACTs, \citealt{Perkins2006, Aleksic2010, Aleksic2012, Arlen2012}) for a long time.
Recently, extended GeV $\gamma$-ray emission from the direction of the Coma cluster has been reported \citep{Xi2018, Adam2021,Baghmanyan2021}.
The former estimation suggested that NGC 4869 and NGC 4874, which are two brightest radio galaxies in the radio band of the cluster, can not account for the entire observed $\gamma$-ray emission\citep{Ackermann2016, Baghmanyan2021}.
However, it still remains uncertain whether the emission is from the diffuse CRs in the ICM or a combination of several unresolved sources in the region.
Under the assumption that the $\gamma$-ray signal comes entirely from the decay of $\pi_0$ produced in $pp$ collisions, the observed $\gamma$-ray flux can set an upper limit for the a CR content in galaxy cluster.
In our previous work \citep{Shi2022}, we studied propagation of CRs in Coma cluster and calculate the radial distribution of generated pionic $\gamma$-rays emission.
By comparing the $\gamma$-ray flux and upper limits obtained by Fermi-LAT and VERITAS for the Coma cluster, we have established an upper limit on the average baryon loading factor for AGNs in the cluster as $\eta_{p,\mathrm{grav}} = W_p/W_\mathrm{grav} \lesssim 0.1$ (or $\eta_{p,\mathrm{rad}} \lesssim 1$). This limit is found to be lower than the baryon loading factor required for blazars as obtained by \citet{Murase2014}.

However, we note that the constraint on the baryon loading factor of AGNs obtained from Coma cluster may not be generalized to the entire AGN population in the universe. A more representative constraint would be based on the all-sky diffuse neutrino flux or all-sky gamma-ray flux. Recently, \citet{Abbasi2022} performed stacking analysis of 1094 galaxy clusters from $Planck$ using 9.5 years of muon-neutrino track events and found no evidence for significant neutrino emission. 
%This results in an upper limit of high-energy diffuse neutrino flux from massive galaxy clusters. 
The differential upper limits presented by IceCube in \citet{Abbasi2022} are most constraining in the energy range between 10 TeV and 1 PeV, suggesting that the contribution of the galaxy cluster population cannot exceed 9\%–13\% of the diffuse neutrino flux.
We may calculate the expected flux of diffuse neutrinos produced by CRs escaping from AGNs via $pp$ collisions in ICM. By comparing the expected flux with the measured upper limit, a constraint on the average baryon loading factor of AGN jets or outflows can be obtained.

The rest of this paper is organized as follows.
In Section \ref{sec:model}, we first review propagation of CRs escaped from AGN in the galaxy cluster, and then calculate the high-energy neutrino production from galaxy clusters through interactions between these CRs and ICM.
In Section \ref{sec:result}, we compare the expected diffuse neutrino flux with the upper limit given by IceCube to constrain the amount of injected protons from AGN, which can be translated to the baryon loading factor.
We summarize and discuss our results in Section \ref{sec:dis}.

\section{Model}
\label{sec:model}
In this section, we describe the model we used in our work. 
In Section 2.1, we review particle diffusion and examine the confinement of CRs in the turbulent magnetic field of galaxy clusters. 
Then, in Section 2.2, we calculate the neutrino emissivity from an individual galaxy cluster. 
Finally, in Section 2.3, we integrate our results and obtain the diffuse neutrino flux contributed by AGNs in galaxy cluster populations.

\subsection{cosmic ray diffusion}
\label{sec:diff}
After acceleration, particles propagate through the turbulent magnetic field of the cluster.
Their diffusion depends on both the particle's Larmor radius $r_\mathrm{L}$ and the coherence length $l_\mathrm{c}$ of the magnetic field.
For typical parameters in the galaxy clusters, when $r_\mathrm{L} < l_\mathrm{c}$, corresponding to the particle's energy $E_p \lesssim 5 \times 10^{20}\ Z (B/5 \ \mathrm{\mu G})(l_\mathrm{c}/0.1\ \mathrm{Mpc})\ \mathrm{eV}$, the propagation of the CRs enters the diffusive regime with the diffusion coefficient of
\begin{equation}
D_\mathrm{cl}
= \frac{1}{3} \left( \frac{B}{\delta B} \right)^2 c r_\mathrm{L}^{2-w} l_\mathrm{c}^{w-1} 
\approx 6.9 \times 10^{31}
\left( \frac{l_\mathrm{c}}{0.1 r_\mathrm{vir,200}} \right)^{2/3}
\left( \frac{M_\mathrm{vir}}{10^{15}M_\odot} \right)^{2/9}
\left( \frac{E_p}{1\ \mathrm{PeV}} \right)^{1/3}  \left( \frac{B Z}{5 \ \mathrm{\mu G}} \right)^{-1/3}
\ \mathrm{cm^2 s^{-1}}.
\end{equation}
For Kolmogorov diffusion, the spectral index is $w=5/3$, we assumed that $B \sim \delta B$ and $l_c \sim$ 10\% of the virial radius $r_\mathrm{vir,200}$ is the typical magnetic field coherence length in the galaxy clusters.
The virial radius $r_\mathrm{vir,200}$ of the cluster with mass $M_\mathrm{vir}$ is defined as $r_\mathrm{vir,200} = (3M_\mathrm{vir}/(4 \pi \Delta_\mathrm{c} \rho_m))^{1/3}$, where $\Delta_\mathrm{c} = 200$.

The diffusion timescale of the CRs in a galaxy cluster with mass $M_\mathrm{vir} = 10^{15} M_\odot$ can be estimated as
\begin{equation}\label{eq:difftime}
t_\mathrm{diff} \approx
\frac{r_\mathrm{vir}^2}{2D_\mathrm{cl}} \approx 11
\left( \frac{M_\mathrm{vir}}{10^{15} M_\odot} \right)^{4/9}
\left( \frac{E_p}{1\ \mathrm{PeV}} \right)^{-1/3}
\left( \frac{BZ}{5\ \mathrm{\mu G}} \right)^{1/3} \ \mathrm{Gyr}.
\end{equation}
When diffusion timescale is longer than the Hubble time $t_\mathrm{H} \sim 14\ \mathrm{Gyr}$, the CRs with energy $E_p \lesssim 0.9 \ \mathrm{PeV}$ are confined in the cluster by magnetic fields.

Then, we calculate the radial density distribution of the CRs at present in a typical galaxy cluster.
As different CR injection histories have slight effect on the result\citep{Shi2022}, we assume a time independent CR injection rate with single power law spectrum of injection index $\alpha$, $dN_p/dE_p \propto E_p^{-\alpha}$.
Neglecting the energy loss of particles, the radial density distribution of the CRs at present can be written as
\begin{equation}
\label{eq:n_p}
n_p(E_p, r, M) \propto \frac{\eta_{p,\rm grav}}{8 \pi^{3/2}}\int_0^{10\ \mathrm{Gyr}}   \frac{E_p^{-\alpha} e^{-r^2/\left(4 D_\mathrm{cl}(E_p,M)t \right)}}{\left( D_\mathrm{cl}(E_p,M)t \right)^{3/2}} dt.
%= \frac{dN_p}{dE_p dV}
\end{equation}
We assume that injection occurs at the center of the galaxy cluster ($r = 0$) and that the injection duration lasts for $\sim$ 10 Gyr.

\subsection{neutrino production}
\label{sec:nu}
As CRs propagate through the cluster magnetic field, they interact with the ICMs and produce $\gamma$-ray photons and neutrinos. 
In this section, we calculate the neutrino flux from a
typical galaxy cluster. 
The hot ICMs emit X-rays via bremsstrahlung radiation, with the emissivity proportional to the square of the number density of electrons in the gas ($n_{\mathrm{ICM},e}$).
Therefore, the density of ICM can be derived from X-ray observations.
From stacking the Chandra data of 320 clusters, the mean density profile of electrons in the ICM can be approximated with a form introduced by \cite{Patej2015}
\begin{equation}
\left( \frac{H(z)}{H_0} \right)^{-2} n_{\mathrm{ICM},e}(x) =
0.00557 \left(\frac{x}{0.201}\right)^{-0.150}
\left(\frac{x}{0.265}\right)^{-0.0638} \left[1+0.59(\frac{x}{0.201})^{0.949}\right]^{-2.936},
\label{eq:ICM}
\end{equation}
where $x=r/r_\mathrm{vir,200}$, $r_\mathrm{vir} = (3M_\mathrm{vir}/(4 \pi \Delta_\mathrm{c} \rho_\mathrm{crit}))^{1/3}$ with $\Delta_\mathrm{c} = 200$, $\rho_\mathrm{crit}=3H^2(z)/(8 \pi G)$. $H(z)$ is the Hubble parameter at redshift $z$, today's Hubble parameter is referred to as the Hubble constant, $H_0$.
In a fully ionized gas, the number density ratio of electron and proton is $n_{\mathrm{ICM},e} = 1.17n_{\mathrm{ICM},p}$.

Following the calculation in \cite{Kelner2006}, we first calculate the $\gamma$-ray emissivity as
\begin{equation}
\label{eq:phi_nu}
\phi_\gamma(E_\gamma, {\bf r}, M,z)\equiv
\frac{dN_\gamma}{dE_\gamma dV dt} \nonumber=
c n_{\mathrm{ICM},p}({\bf r},M,z)
\int_{E_\gamma}^{\infty} \sigma_{pp}(E_p) n_p(E_p, r, M) 
 F_\gamma(\frac{E_\gamma}{E_p}, E_p) \frac{dE_p}{E_p},
\end{equation}
where $\sigma_{pp} (E_p)$ is the total inelastic cross section of $pp$ interactions, $F_\gamma (E_\gamma/E_p, E_p)$ is the spectrum of the secondary $\gamma$-ray in a single collision.

Due to the advantage of the angular resolution of muon-track events, source analyses presented with IceCube usually focus on (anti-)muon neutrinos.
Assuming equal amount of neutrinos of three flavors after the oscillation, we can relate the muon neutrino emissivity to that of $\gamma-$rays as
\begin{equation}
    E_\gamma^2 \phi_{E_\gamma} \approx 2 E_\nu^2 \phi_{E_\nu} |_{E_\nu = E_\gamma/2}
\end{equation}
Integrating the total extent of the galaxy cluster, the total muon neutrino flux from an individual galaxy cluster can be calculated as
\begin{equation}
E_\nu^2 \Phi_0(E_\nu, M, z) =
\frac{E_\nu^2 dN_\nu}{dE_\nu dt} =
%\frac{1}{2 \pi (1- \mathrm{cos} \psi_\mathrm{vir})}
\int_0^{\psi} \frac{E_\nu^2 \phi_\nu(E_\nu, {\bf r})}{4 \pi {d_\mathrm{L}}^2 (z)} dV,
\end{equation}
where $dV = r^2 \mathrm{sin} \theta dr d\theta d\varphi$ is the differential volume element of coordinate {\bf r}$=(r, \theta, \varphi)$,
%$2 \pi(1- \mathrm{cos} \psi_\mathrm{vir})$ is the solid angle of the neutrino flux,
$\psi$ is the angular extension of the source, $d_\mathrm{L}$ is the luminosity distance of the cluster.

\begin{figure}[ht]
    \centering
    \includegraphics[width=0.49\textwidth]{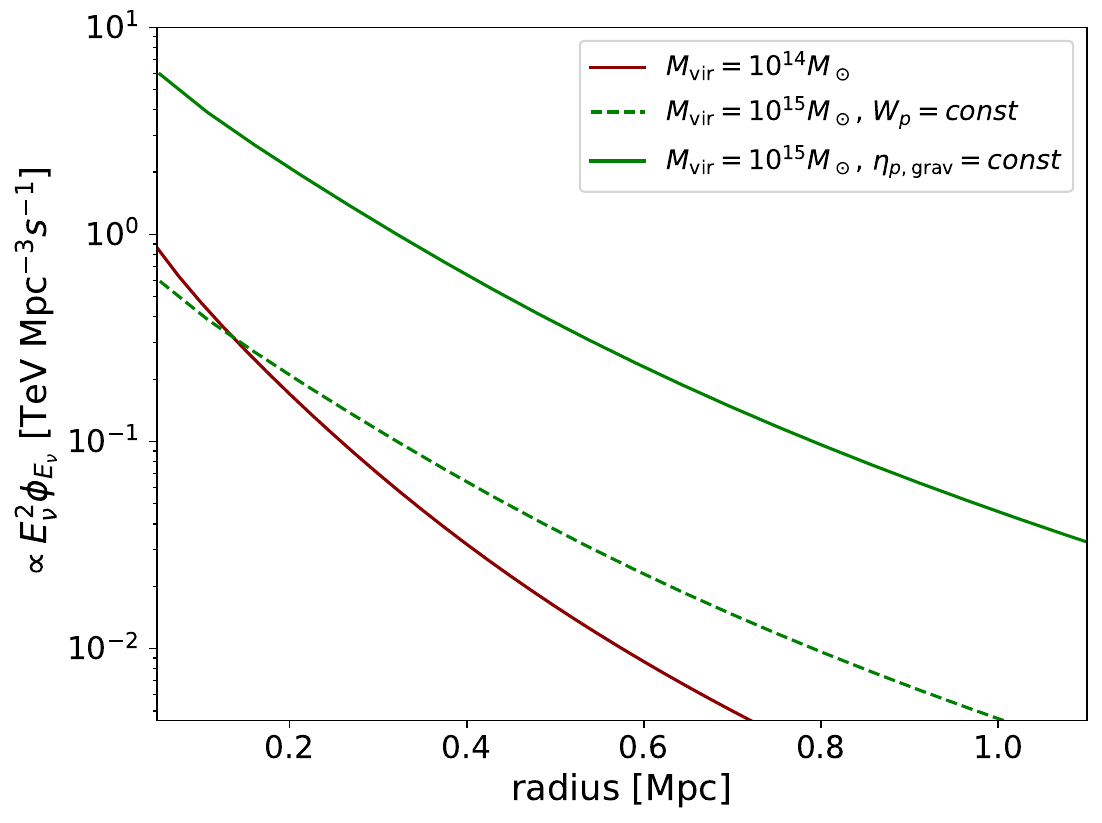}
    \includegraphics[width=0.48\textwidth]{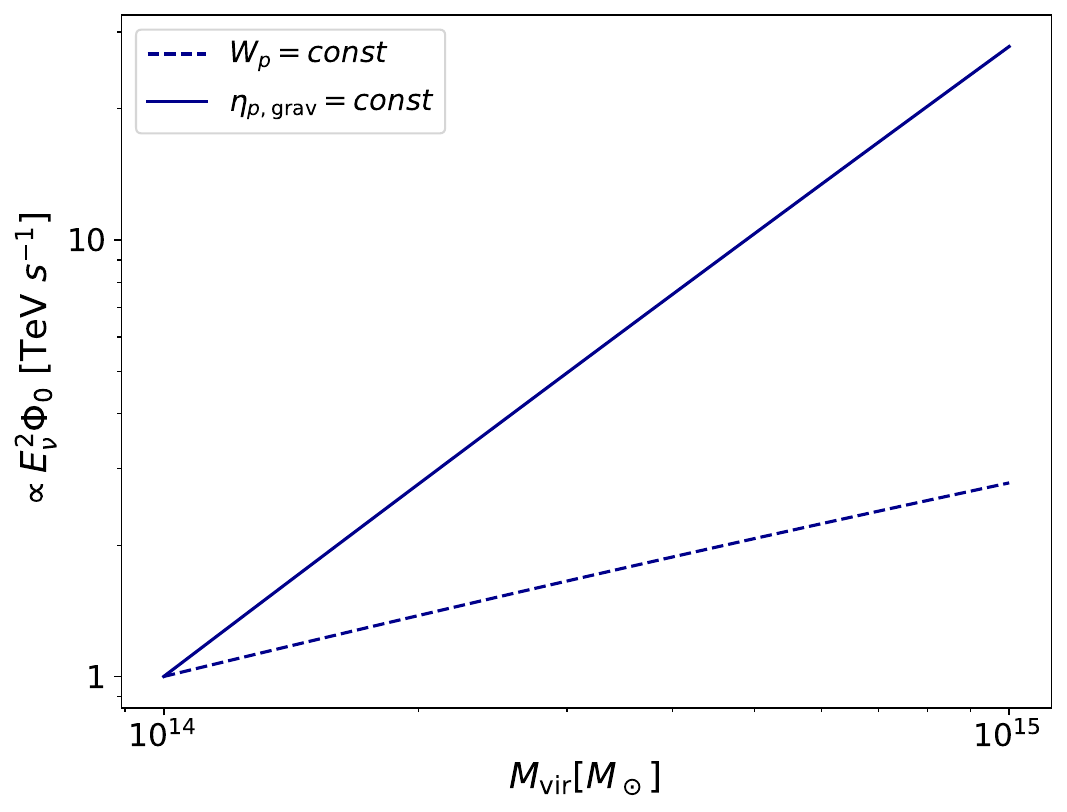}
    \caption{
    {\bf Left:} The normalized radial density distribution of neutrino emissivity with energy $E_\nu = 1$ TeV for a cluster with viral mass $M_\mathrm{vir} = 10^{14} M_\odot$ (red), $10^{15} M_\odot$ (green) and redshift $z=0$.
    The flux is normalized by fixing it to unity at the center of the cluster with $M_\mathrm{vir} = 10^{14} M_\odot$.
    The total neutrino emissivity derived from $l_\mathrm{c} = 0.1 r_\mathrm{vir,200}$ under different CR injection rates, regarding either a same total CR injection $W_p = const$ (dashed line) or a constant baryon loading factor $\eta_{p,\rm grav} = const$ (solid line). The latter is the model employed in our study.
    {\bf Right:} The relation between the viral mass of a galaxy cluster and the total neutrino emissivity at redshift $z=0$. The emissivity is normalized by fixing the value to be unity for the cluster of $M_{\rm vir} = 10^{14} M_\odot$. The line types correspond to the same assumptions as in the left panel. }
    \label{fig:nu_M}
\end{figure}

The virial mass of a galaxy cluster can affect the neutrino emissivity in our model. To study this influence, we compare 
the normalized radial distribution of the neutrino emissivity at $E_\nu = 1$ TeV for cluster of different viral masses.
First, the energy budget of CRs in a more massive cluster is higher if the same fraction of gravitational energy is converted to cosmic rays (i.e., the same $\eta_{p,\rm, grav}$). Also, a more massive cluster is surrounded by a larger amount of ICM, leading to a higher gas density compared at the same radius $r$. It results in a higher interaction rate of $pp$ collisions and consequently increase the neutrino emissivity. On the other hand, the magnetic field coherence length $l_c$ is related with the virial mass by $l_c=0.1r_{\rm vir}\propto M_{\rm vir}^{1/3}$. The resulting diffusion coefficient is positively related to the virial mass and hence particles diffuse faster in more massive clusters, leading to a flatter radial distribution of the neutrino emissivity in a more massive cluster than in a less massive one. Note that the virial radius increases with the virial mass as $M_{\rm vir}^{1/3}$, and hence the diffusive escape timescale increases with the virial mass as shown in Eq.~\ref{eq:difftime}. Therefore, even with the same CR injection rate, the total neutrino luminosity still increases with virial mass. The comparison of the radial distribution of the neutrino emissivity between a $10^{14}M_\odot$ cluster and a $10^{15}M_\odot$ cluster is shown in the left panel of Figure~\ref{fig:nu_M}. A direct dependence of the total neutrino luminosity on the virial mass of the cluster is shown in the right panel of Figure~\ref{fig:nu_M}. We observed a monotonic increase of the neutrino luminosity with the virial mass, approximately with a linear relation.

To ensure consistency with the results from \cite{Abbasi2022}, we integrate the total neutrino flux over the extent of the cluster from 0 to the virial mass $r_\mathrm{vir,500}$, where $\Delta_\mathrm{c} = 500$, when calculating the total neutrino flux. 
As the neutrino flux is concentrated in the central region of the galaxy cluster, the integration range have little effect on the final results.

\subsection{diffuse neutrino flux}
\label{sec:nu_diff}
The diffuse muon neutrino flux integrated from the entire galaxy clusters can be estimated as
\begin{equation}
    E_\nu^2 \Phi_\nu(E_\nu) = \int dlnM \frac{dn}{dlnM} (1+z)^2 E_\nu^2 \Phi_0 \frac{dV_c}{d\Omega},
\end{equation}
where the differential number density $dn/dM$ of clusters with mass $M$ at redshift $z$ can be obtained from the halo mass function
\begin{equation}
\frac{dn}{dM}(M,z) =
f(\sigma)\frac{\rho_m}{M}
\frac{d \mathrm{ln} \sigma}{dM},
\end{equation}
and $\rho_m$ is the mean density of the universe at the epoch of analysis, $\rho_m (z) = \Omega_m (z) \rho_\mathrm{crit} (z) = \rho_m (0)(1+z)^3$,
$\sigma(M,z)$ is the rms variance of the linear density field smoothed on scale $R= (3M/4 \pi \rho_m)^{1/3}$,
and $f(\sigma)$ describes the $\sigma$-weighted distribution.
$dV_c = c(1+z)^2 /H(z)d_A^2dz d\Omega$ is the co-moving volume and $d_A = d_L (1+z)^{-2}$ is the angular diameter distance.

In this study, we adopted the same mass halo function from \cite{Tinker2010} as used in \cite{Abbasi2022} to ensure consistency.
As galaxy clusters with masses below $10^{14} M_\odot$ or $z >1$ are not expected to produce a significant flux of neutrinos at earth\citep{Fang2016},
our calculation only considers clusters with masses between $10^{14} M_\odot$ and $10^{15} M_\odot$ and a redshift between 0.01 and 2.

\section{Constraining the Baryon Loading Factor}
\label{sec:result}
In this section, we calculate the average baryon loading factor $\eta_{p,\rm grav}$ for the population of galaxy clusters.
Although the value of $\eta_{p,\rm grav}$ may vary for each cluster, our primary concern lies in the constraints on the total population of galaxy clusters. Therefore, we focus on calculating the average value of $\eta_{p,\rm grav}$ for the entire cluster population.
We use the same definition of baryon loading factor $\eta_{p,\mathrm{grav}}$ in Section 4.1 from \citet{Shi2022} to constrain the total efficiency of releasing gravitational potential energy $W_g$ loading into total injected energy of the baryons $W_p$, written as
\begin{equation}
    \eta_{p,\mathrm{grav}} = W_p/W_g,
\end{equation}
where the total injected energy of the baryons $W_p$ can be calculated from the total CR injection rate $W_p = \int E_p (dN_p/dE_p) dE_p dt$.
The releasing gravitational potential energy $W_g$ is estimated as $W_g = 0.2 M_\mathrm{BH} c^2$ with an intermediate mass-to-energy conversion efficiency between standard accretion disk model $\approx 0.1$\citep{Shakura1976} and a extreme Kerr black hole $\approx 0.3$\citep{Thorne1974}.
The total black hole mass $M_\mathrm{BH}$ in the cluster can be estimated as a fraction to the viral mass $M_\mathrm{vir}$, $M_\mathrm{BH} = \eta_\mathrm{BH} M_\mathrm{vir}$.
Noted that, this fraction is linearly correlated to the total gravitational energy $W_g$ and hence would linearly affect the obtained value of $\eta_{p,\rm grav}$.
For the Coma cluster, the galaxy mass can be integrated with the help of the mass to (V-band) light ratio, $M_\mathrm{gal} = 2.03 \times 10^{13} M_\odot$ while the viral mass is measured to be $M_\mathrm{vir,500} = 6.0 \times 10^{14} M_\odot$. Therefore, the ratio of galaxy mass to the viral mass is fixed at $\eta_\mathrm{gal} = M_\mathrm{gal}/M_\mathrm{vir} = 0.034$ in our study, which is also consistent with the standard $\Lambda$CDM model.
The fraction of total black hole mass to the galaxy mass can be estimated as $\eta_\mathrm{BH,gal}= 0.002–0.006$ \citep{Kormendy1995,Wang1998}.
To be conservative, we estimated $\eta_\mathrm{BH} = \eta_\mathrm{BH,gal} \eta_\mathrm{gal} \approx 0.002 \times 0.034 = 6.8 \times 10^{-5}$.

\begin{figure}
    \centering
    \includegraphics[width=\textwidth]{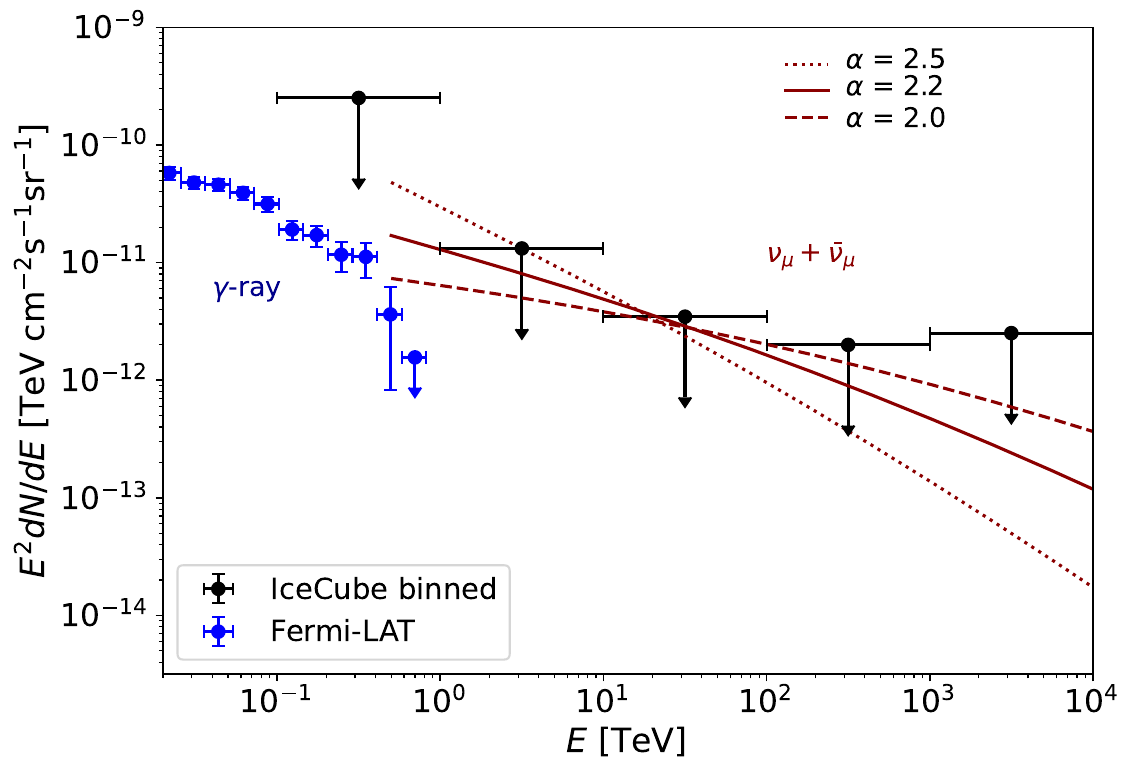}
    \caption{
     Integrated muon neutrino flux (red lines) from galaxy clusters in this work.
     Different injection spectral index $\alpha =$2(dashed), 2.2(solid), 2.5(dotted) are illustrated separately.
     It is compared with the differential upper limits in one-decade energy bins for the distance weighting ($1/d^2$) scheme from IceCube analysis\citep{Abbasi2022}.
     The IGRB observed by Fermi-LAT\citep{Ackermann2015} is also shown for comparison.
     }
    \label{fig:sed_nu}
\end{figure}

\citet{Abbasi2022} obtained the neutrino flux upper limit from the galaxy cluster population for on two different weighting methods for the expected neutrino flux from a galaxy cluster. One is the so-called distance weighting, assuming neutrino luminosity is the same among all clusters and hence the neutrino flux is proportional to $1/d_L^2$. The other is the mass weighting, assuming the neutrino luminosity scales linearly with the virial mass of the cluster and hence the neutrino flux from an individual cluster scales with $M_{\rm vir}/d_L^2$. According to our discussion in the previous section and the result shown in Fig.~\ref{fig:nu_M}. The mass weighting is more consistent with our model. However, in the analysis, they fixed the neutrino spectrum to be an unbroken power-law function with a slope of $-2.5$. As we want to test different values of the spectral index $\alpha$ for the injected protons, it is not appropriate to directly compare our results with the flux upper limit derived with the mass weighting method. On the other hand, \citet{Abbasi2022} also provided the quasi-differential flux upper limits of 90\% confidence level in one-decade energy bins with the distance weighting, we choose to constrain the value of $\eta_{p,\rm grav}$ by comparing our results with this differential upper limit. Note that the neutrino flux upper limit obtained with the mass weighting is stricter than that obtained with the distance weighting. So, such a comparison leads to a conservative constraints on the baryon loading factor $\eta_{p,\rm grav}$.

\begin{figure}
    \centering
    \includegraphics{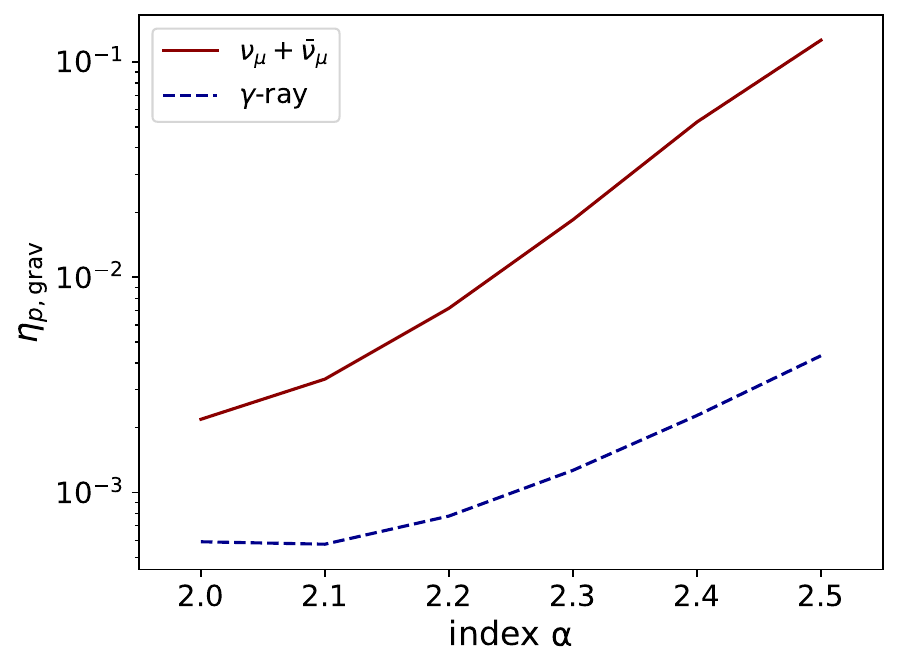}
    \caption{The upper limits of baryon loading factor $\eta_{p,\rm grav}$ constrained by diffuse muon neutrino and $\gamma$-ray flux with different indices $\alpha$ of the injected protons from 2.0 to 2.5.
}
    \label{fig:etap}
\end{figure}

Using the same cluster mass ranging from $10^{14} M_\odot$ to $10^{15} M_\odot$ and the redshift between 0.01 and 2, we calculate the results and compare with the differential upper limits from \citet{Abbasi2022} in Figure~\ref{fig:sed_nu}.
%The comparison is shown in Figure~\ref{fig:sed_nu}.
Since the modeled neutrino flux linearly depends on $\eta_{p,\rm grav}$, we can find out the maximally allowed value of $\eta_{p,\rm grav}$ that makes the modeled neutrino flux saturating only one of the five bins for the flux upper limit. For reference, we also display the isotropic diffuse $\gamma$-ray background (IGRB) observed by Fermi-LAT\citep{Ackermann2015} in the figure. Furthermore, we explore the maximum $\eta_{p,\rm grav}$ with different CR spectral injection index $\alpha$, and obtain the relation between $\alpha$ and the upper limit of $\eta_{p,\rm grav}$ as shown in Figure \ref{fig:etap}.
For a flat injection CR spectrum $\alpha$ = 2, which is expected under the canonical shock acceleration theory, a quite strict constraint $\eta_{p,\rm grav} = 2 \times 10^{-3}$ can be obtained. The constraint becomes less stringent for a softer spectral index, and the upper limit of $\eta_{p,\rm grav}$ increases up to 0.12 for $\alpha = 2.5$.

We note that the diffuse neutrino flux from galaxy cluster highly depends on model parameters such as the baryon loading factor $\eta_{p,\rm grav}$ and others. In previous literature \citep{Fang2016, Fang2018, Hussain2022}, the authors aim to explore the potential of galaxy clusters as the major sources of the all-sky diffuse high-energy neutrino flux, so they tune the model parameters to make the predicted neutrino flux match the measured one. The latest observational upper limits of the diffuse neutrino flux from galaxy clusters presented by IceCube in \citet{Abbasi2022} suggest that the contribution of the galaxy cluster population cannot exceed 9\%–13\% of the diffuse flux, which can thus be also used to constrain the model parameters in previous literature, as we do in the present study.

\section{discussion}
\label{sec:dis}

\subsection{cumulative $\gamma$-ray flux}

The CR-ICM interaction can also produce $\gamma$-ray photons via the decay of $\pi^0$s.
There are suggestions that the observed diffuse neutrino/$\gamma$-ray flux could be completely explained by the cumulative emission from galaxy clusters \citep{Hussain2022, Fang2018}.
However, the origin of the IGRB is still under-debate, besides the galaxy clusters, the observed IGRB is possibly superimposed by different populations of gamma-ray emitters such as star-forming galaxies, starburst galaxies and active galactic nuclei.

We calculate the diffuse $\gamma$-ray flux from the clusters of galaxies in the same way as for neutrinos, taking into account the effect of extragalactic background light (EBL) attenuation using the model from \cite{Saldana-Lopez2021}.
The dominant contribution to the total flux of $\gamma$-rays comes from sources at low redshifts ($z\lesssim 0.3$), where the effect of EBL attenuation is less pronounced.

\begin{figure}
    \centering
    \includegraphics[width=\textwidth]{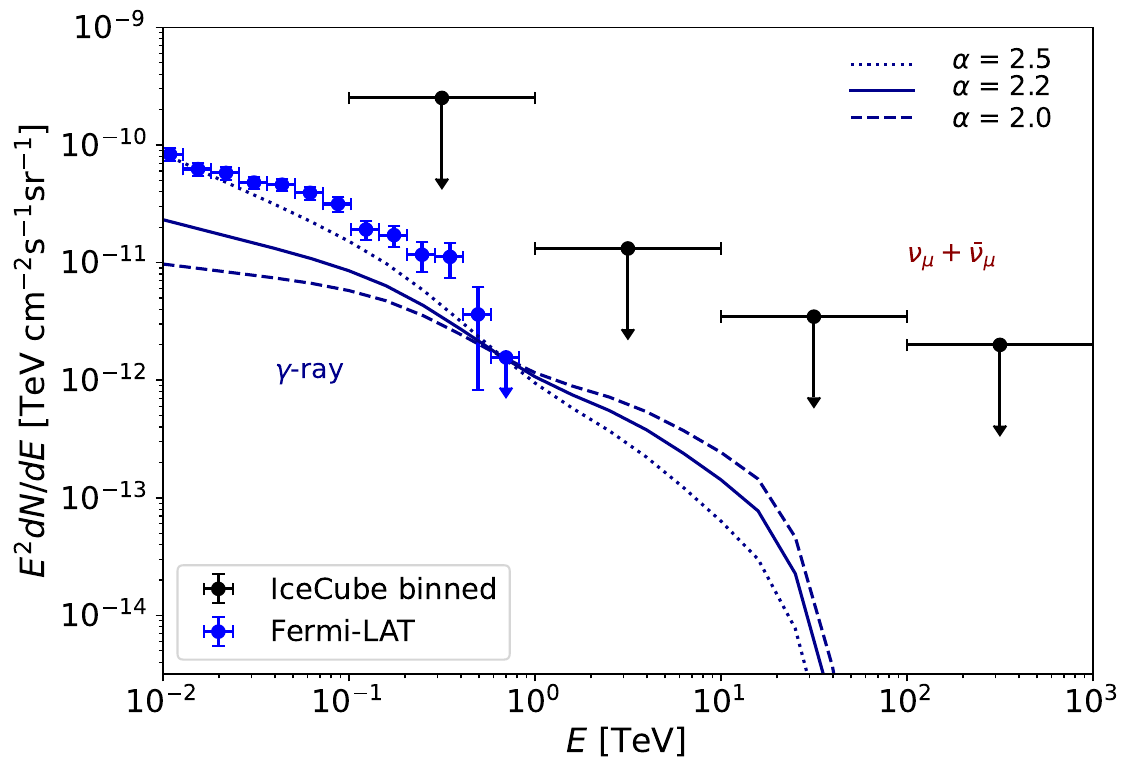}
    \caption{Cumulative $\gamma$-ray flux from the galaxy clusters for different proton injection spectral index $\alpha$ = 2.0 (dashed), 2.2 (solid), 2.5 (dotted). The flux is compared with the IGRB observations from Fermi-LAT. The differential upper limits for the distance weighting ($1/d^2$) scheme from IceCube analysis are also shown in the figure for comparison.
}
    \label{fig:sed_gamma}
\end{figure}

Considering the IGRB observed by Fermi-LAT as an upper limit, we can also derive the upper limit of the baryon loading factor from the integrated $\gamma$-ray flux.
Figure \ref{fig:sed_gamma} compares the diffuse flux of $\gamma$-ray from the clusters of galaxies and the IGRB measured by Fermi-LAT. The upper limits of neutrino flux obtained by IceCube is also shown for reference.
For considered injection indices, i.e., $2\leq \alpha\leq 2.5$, the constraints derived from the diffuse $\gamma$-ray flux are 1-2 orders of magnitude more stringent than those from neutrinos.

However, there may be some potential uncertainties in the constraints obtained from the diffuse $\gamma$-ray flux. This is mainly because the corresponding energies of CRs responsible for these $\gamma$-ray emission are relatively low. \citet{Shi2022} suggested that the propagation of these CRs may be influenced by the streaming instability \citep{Kulsrud1969, Skilling1971}. As a result,  the CR spatial distribution may be different from what is predicted in our current model and the CR energies may be dissipated through self-excited Alfv{\'e}n waves in the ICM. In addition, these relatively low-energy CRs may not be so easy to escape their acceleration sites and could lose energy adiabatically \citep{Fang2018}. These processes are not considered in our model.
On the other hand, we should also note that the most constraining energy bin of the Fermi-LAT data is the highest-energy one in [580, 820]\,GeV (as shown in Figure \ref{fig:sed_gamma}), corresponding to the CR proton energy of $\lesssim 10\,$TeV. As a result, the aforementioned uncertainties may not be so severe. Another uncertainty arise from the accuracy of the EBL model, which may affect the attenuated $\gamma$-ray spectrum \citep{Hussain2022}.

\subsection{distinguish from the acceleration caused by cluster mergers}
In galaxy clusters, CRs can also be accelerated by shock waves arising from cluster merger processes.
These shocks would also accelerate particles to relativistic energies and produce $\gamma$-ray and neutrino emission \citep{Colafrancesco1998,Ryu2003}.
The gravitational energy released of two galaxy clusters with total mass of $10^{15} M_\odot$ merging from a infinity distance to 1\,Mpc is $\sim 10^{64}$\,erg. Most of this gravitational energy is converted into the kinetic energy of dark matter, only about 10\% is dissipated into the ICM.
The gravitational energy released from central black hole accretions in our study is estimated as $\sim 10^{64}$ erg for a galaxy cluster with viral mass $10^{15} M_\odot$.
Due to the lacking knowledge of the baryon loading efficiency of these two acceleration mechanisms, it is hard to tell which process dominates CR accelerations in the galaxy cluster.

The future observed $\gamma$-ray morphology may help us distinguish which process would dominate the acceleration.
In the central AGN injection model, the $\gamma$-ray profile is a halo-like structure which clusters in the center region and decreases with radius.
Numerical simulations modelling the formation of large-scale structures shows the presence of strong accretion shocks in the outer regions of galaxy clusters\citep{Miniati2000,Ryu2003,Vazza2012}, the radial density profile of CRs and $\gamma$-ray predicted by the cluster mergers would be a shell-like structure and significantly different from the central injection model.

Studies have shown that the CR-to-thermal pressure ratio from the injection of CRs at cosmological shocks should slightly increase with radius \citep{Vazza2012}. 
However, in our model considering the CR injection from central AGNs, this ratio decreases with the radius to the cluster center unless the energy of the injected proton exceeds $\sim$1 PeV.
The CR-to-thermal energy ratio has been limited to $\lesssim$4-10\% for photon indices $\alpha = 2.0-3.2$ using stacked Fermi-LAT count maps in the $1-300$\,GeV band \citep{Huber2013}.
In our work, the upper limit on the average CR-to-thermal pressure ratio for proton indices of $2.0-2.5$ (corresponding to a photon index of $\sim 2.2-2.7$) from AGN contributions is estimated to be $0.3\%-1.2\%$, which is more stringent than (and consistent with) the limitation obtained from stacking Fermi-LAT count maps.

Moreover, the radial density profile of CRs accelerated by AGNs is clustered in the central region of the galaxy cluster, where the gas density of ICMs is higher. 
This enhances the overall interaction rates of the $pp$ collision and the neutrino/$\gamma$-ray production efficiency.
It should be noted that our main purpose is to constrain the upper limit of the baryon loading efficiency. 
Therefore, considering the $\gamma$-ray emission additionally contributed by other processes would not invalidate the results, but would only make the constraints tighter.

\section{conclusions}
\label{sec:con}
In this work, we calculated the diffuse neutrino and $\gamma$-ray flux from the galaxy cluster population and compared the results with the upper limit using 9.5 years of muon-track IceCube data and the IGRB observed by Fermi-LAT. 
In order to compare with the upper limits of the diffuse neutrino flux presented by \citet{Abbasi2022}, the same mass range of galaxy clusters from  $10^{14} M_\odot$ to $10^{15} M_\odot$ and the redshift between 0.01 and 2 are considered in our calculations.
Our best constraint for the upper limits of the average baryon loading factor is $\eta_{p,\rm grav} \lesssim 2 \times 10^{-3}$, assuming a flat injection spectrum index of CRs ($\alpha = 2$). 
When varying the injected power-law spectrum index from 2 to 2.5, we derive the upper limit of the average baryon loading factor as $\eta_{p,\rm grav} \lesssim 0.1$.
The constraints using the IGRB observed by Fermi-LAT are about 1-2 orders of magnitude stricter than the constraints derived from the diffuse neutrino flux from the galaxy cluster population, assuming that the cumulative gamma-ray flux from clusters is the dominant component of the IGRB. We note that the constraint from IGRB is based on propagation model of lower energy (a few TeV) CRs in the ICM. There may be additional physical effects on these lower energy CRs and hence the obtained upper limits may not be valid. 
The constraints derived from both the upper limit on cumulative neutrino flux by IceCube analysis and the IGRB observed by Fermi-LAT are more robust than the one inferred from the $\gamma$-ray observations of the Coma cluster in \citet{Shi2022}.
Finally, we should bear in mind that the constraints on the baryon loading factor obtained here is valid for the entire source population over a long period of time comparable to the Hubble timescale.  It is possible for an individual galaxy cluster or AGNs to exceed this limit, in particular during a short period of time such as AGN flares.

\begin{acknowledgments}
This work is supported by National Natural Scientific Foundation of China under grants. No.U2031105, 12121003, and 12203022,  the National Key R\&D Program of China under grant No. 2022YFF0711404, China Manned Space Project (CMS-CSST-2021-B11). X.S. is supported by China Postdoctoral Science Grant (No. 2022T150314).
\end{acknowledgments}

%\appendix
%\section{Appendix information}

\bibliographystyle{aasjournal}

\end{document}